\documentclass[iop,revtex4]{emulateapj}
\usepackage{graphicx,color}
\usepackage{amssymb}
\usepackage{amsmath}
\usepackage{threeparttable}
\usepackage{lscape}
\usepackage{longtable}
\usepackage{changepage}
\usepackage{hyperref}
\hypersetup{colorlinks   = true, citecolor    = blue}

\mathchardef\mhyphen="2D

\newcommand{\oiii}{[O\,{\sc iii}]}
\newcommand{\ovi}{O\,{\sc vi}}

\newcommand{\siv}{S\,{\sc iv}}
\newcommand{\siiv}{Si\,{\sc iv}}

\newcommand{\pv}{P\,{\sc v}}
\newcommand{\cii}{[C\,{\sc ii}]}
\newcommand{\ciii}{C\,{\sc iii}}
\newcommand{\civ}{C\,{\sc iv}}
\newcommand{\mgii}{Mg\,{\sc ii}}

\newcommand{\angstrom}{\text{ \normalfont\AA}}

\newcommand{\sub}[2]{\ifmmode #1_\mathrm{\scriptstyle #2} \else $#1_\mathrm{\scriptstyle #2}$\fi}
\newcommand{\ssub}[2]{\ifmmode #1_\mathrm{\scriptscriptstyle #2} \else $#1_\mathrm{\scriptscriptstyle #2}$\fi}

\mathchardef\mhyphen="2D

\definecolor{blk}{rgb}{0.0,0.0,0.0}
\definecolor{red}{rgb}{0.75,0.0,0.0}
\definecolor{yel}{rgb}{0.65,0.65,0.0}
\definecolor{grn}{rgb}{0.0,0.75,0.0}
\definecolor{blu}{rgb}{0.0,0.0,0.75}
\definecolor{gry}{rgb}{0.75,0.75,0.75}

\def\lya{Ly$\alpha$}
\def\ly{$\lambda$}
\def\hi{H\,{\sc i}}

\def\cii{C\,{\sc ii}}
\def\ciii{C\,{\sc iii}}
\def\civ{C\,{\sc iv}}
\def\niii{N\,{\sc iii}}
\def\niv{N\,{\sc iv}}
\def\nv{N\,{\sc v}}
\def\oi{O\,{\sc i}}
\def\oiii{O\,{\sc iii}}
\def\oiv{O\,{\sc iv}}
\def\ov{O\,{\sc v}}
\def\ovi{O\,{\sc vi}}

\def\neviii{Ne\,{\sc viii}}

\def\naix{Na\,{\sc ix}}

\def\mgx{Mg\,{\sc x}}
\def\mgii{Mg\,{\sc ii}}

\def\siiv{Si\,{\sc iv}}

\def\pv{P\,{\sc v}}
\def\siv{S\,{\sc iv}}

\def\svi{S\,{\sc vi}}
\def\ariv{Ar\,{\sc iv}}

\def\arviii{Ar\,{\sc viii}}

\def\cavii{Ca\,{\sc vii}}
\def\caviii{Ca\,{\sc viii}}

\def\HP{VHP}
\def\LP{HP}

\def\nh{\ifmmode n_\mathrm{\scriptscriptstyle H} \else $n_\mathrm{\scriptscriptstyle H}$\fi}
\def\ne{\ifmmode n_\mathrm{\scriptstyle e} \else $n_\mathrm{\scriptstyle e}$\fi}
\def\Qh{\ifmmode Q_\mathrm{\scriptstyle H} \else $Q_\mathrm{\scriptstyle H}$\fi}
\def\Uh{\ifmmode U_\mathrm{\scriptstyle H} \else $U_\mathrm{\scriptstyle H}$\fi}
\def\Nh{\ifmmode N_\mathrm{\scriptstyle H} \else $N_\mathrm{\scriptstyle H}$\fi}
\def\Uhhp{\ifmmode U_\mathrm{\scriptstyle H,HP} \else $U_\mathrm{\scriptstyle H,HP}$\fi}
\def\Nhhp{\ifmmode N_\mathrm{\scriptstyle H,HP} \else $N_\mathrm{\scriptstyle H,HP}$\fi}
\def\Uhvhp{\ifmmode U_\mathrm{\scriptstyle H,VHP} \else $U_\mathrm{\scriptstyle H,VHP}$\fi}
\def\Nhvhp{\ifmmode N_\mathrm{\scriptstyle H,VHP} \else $N_\mathrm{\scriptstyle H,VHP}$\fi}
\def\Nion{\ifmmode N_\mathrm{\scriptstyle ion} \else $N_\mathrm{\scriptstyle ion}$\fi}

\def\Zsun{\ifmmode {\rm Z}_{\odot} \else Z$_{\odot}$\fi}
\def\Msun{\ifmmode {\rm M}_{\odot} \else M$_{\odot}$\fi}
\def\kms{\ifmmode {\rm km~s}^{-1} \else km~s$^{-1}$\fi}
\def\Lya{\ifmmode {\rm Ly}\alpha \else Ly$\alpha$\fi}
\def\Lyb{\ifmmode {\rm Ly}\beta \else Ly$\beta$\fi}
\def\Lyg{\ifmmode {\rm Ly}\gamma \else Ly$\gamma$\fi}
\def\Lyd{\ifmmode {\rm Ly}\delta \else Ly$\delta$\fi}
\def\neaod{\ifmmode n_\mathrm{\scriptscriptstyle AOD} \else $n_\mathrm{\scriptscriptstyle AOD}$\fi}
\def\necrit{\ifmmode n_\mathrm{\scriptstyle cr} \else $n_\mathrm{\scriptstyle cr}$\fi}
\def\ncr{\ifmmode n_\mathrm{\scriptstyle cr} \else $n_\mathrm{\scriptstyle cr}$\fi}
\def\nepi{\ifmmode n_\mathrm{\scriptscriptstyle PI} \else $n_\mathrm{\scriptscriptstyle PI}$\fi}
\def\gtorder{\mathrel{\raise.3ex\hbox{$>$}\mkern-14mu\lower0.6ex\hbox{$\sim$}}}
\def\ltorder{\mathrel{\raise.3ex\hbox{$<$}\mkern-14mu\lower0.6ex\hbox{$\sim$}}}

\newcommand{\Comp}{System}
\newcommand{\comps}{systems}

\slugcomment{Submitted to ApJS 2019 Jul 14; Accepted 2019 Dec 4}

\shorttitle{ }
\shortauthors{Xu et al.}
\shortauthors{}

\begin{document}


\title{HST/COS observations of quasar outflows in the 500 -- 1050\angstrom\ rest frame: \\VI  Wide, Energetic Outflows in SDSS J0755+2306}


\author{
Xinfeng Xu\altaffilmark{1,$\dagger$},
Nahum Arav\altaffilmark{1},
Timothy Miller\altaffilmark{1},
Gerard A. Kriss\altaffilmark{2},
Rachel Plesha\altaffilmark{2}
}

\affil{$^1$Department of Physics, Virginia Tech, Blacksburg, VA 24061, USA\\
$^2$Space Telescope Science Institute, 3700 San Martin Drive, Baltimore, MD 21218, USA\\
\hspace{07mm}\
}

\altaffiltext{$\dagger$}{Email: xinfeng@vt.edu}

\begin{abstract}

We present the analysis of two outflows (S1 at --5500 km s$^{-1}$ and S2 at --9700 km s$^{-1}$) seen in recent HST/COS observations of quasar SDSS J0755+2306 (z = 0.854). The outflows are detected as absorption troughs from both high-ionization species, including \niii, \oiii, and \siv, and very high-ionization species, including \arviii, \neviii, and \naix. The derived photoionization solutions show that each outflow requires a two ionization-phase solution. For S1, troughs from \siv*\ and \siv\ allow us to derive an electron number density, \ne\ = 1.8$\times$10$^4$ cm$^{-3}$, and its distance from the central source of $R$ = 270 pc. For S2, troughs from \oiii*\ and \oiii\ yield \ne\ = 1.2$\times$10$^3$ cm$^{-3}$ and $R$ = 1600 pc. The kinetic luminosity of S2 is $>$ 12\% of the Eddington luminosity for the quasar and therefore can provide strong AGN feedback effects. Comparison of absorption troughs from \oiii\ and \ovi\ in both outflow systems supports the idea that for a given element, higher ionization ions have larger covering fractions than lower ionization ones.

\end{abstract}


\section{INTRODUCTION}
  
Broad absorption line (BAL) outflows are detected as blueshifted absorption troughs in 15 -- 25 \% of quasar spectra \citep[][and references therein]{Tolea02, Hewett03, Reichard03, Trump06, Ganguly08, Gibson09}.  These outflows provide an important mechanism to carry energy, mass, and momentum out of the quasar's central regions \citep[e.g.,][]{Scannapieco04, Ciotti09, Ostriker10, Hopkins10, Choi14, Hopkins16}. Theoretical studies and simulations show that these outflows are related to a variety of AGN feedback processes (see elaboration in section 1 of Arav et al. 2019, hereafter, Paper I).
To quantify the extent that outflows can contribute to AGN feedback, we need to determine their kinetic luminosity ($\dot{E}_{k}$). Theoretical models predict that $\dot{E}_{k}$ needs to be at least 0.5 \% \citep{Hopkins10} or 5~\% \citep{Scannapieco04} of the Eddington luminosity (L$_{edd}$) in order to provide strong AGN feedback. 

In this paper, we analyze two outflows emanating from quasar SDSS J0755+2306. The data is from a spectroscopic survey of ten quasars in the 500 -- 1050\angstrom\ Extreme-UV (EUV500) band (see Paper I). These two outflows present features different from other quasar outflows observed in the EUV500: 1) deep absorption troughs from doubly ionized species, e.g., \ciii\ \ly 977.02\angstrom, the \niii\ multiplets near 686\angstrom, 764\angstrom, and 990\angstrom, and the \oiii\ multiplets near 703\angstrom\ and 834\angstrom; 2) continuous blended absorption that depress the flux in the 1227\angstrom\ $ < \lambda\ < $ 1290\angstrom\ and 1340\angstrom\ $ < \lambda\ < $ 1440\angstrom\ observed frame regions. 

This paper is part of a series of publications describing the results
of  HST program GO-14777, which observed quasar outflows in the EUV500 using the Cosmic Origin Spectrograph (COS).\\
Paper I \citep{ara20a} summarizes the results
for the individual objects and discusses their importance to various
aspects of quasar outflow research. \\
Paper II \citep{xu20a} gives the full
analysis for 4 outflows detected in SDSS J1042+1646, including the
largest kinetic luminosity ($\dot{E}_k$ = $10^{47}$ erg s$^{-1}$) outflow measured to date
at $R=800$~pc, and an outflow at $R=15$~pc. \\
Paper III \citep{mil20a} analyzes 4 outflows
detected in 2MASS J1051+1247, which show remarkable similarities, are
situated at $R\sim200$~pc and have a combined $\dot{E}_k=10^{46}$ erg
s$^{-1}$.  \\
Paper IV \citep{xu20b} presents the largest
velocity shift and acceleration measured to date in a BAL outflow.\\  
Paper V \citep{mil20b} analyzes 2 outflows
detected in PKS 0352-0711, one outflow at $R=500$~pc and a second
outflow at $R=10$~pc that shows an ionization-potential-dependent
velocity shift for troughs from different ions.\\ 
Paper VI is this work.\\
Paper VII \citep{mil20c} discusses the other
objects observed by program GO-14777, whose outflow characteristics
make the analysis more challenging.\\

The structure of this paper is as follows. We present the observations and data reduction in section \ref{sec:data}. In section \ref{sec:spec}, we present the analysis of the spectrum for the two outflow systems. We determine each outflow' electron number density (\ne) and distance in section \ref{sec:distance} and constrain their energetics in section \ref{sec:energy}. We discuss the results and compare with other EUV500 outflows in section \ref{sec:discussion}; and summarize the paper in section \ref{sec:summary}. We adopt a cosmology with H$_{0}$ = 69.6 km s$^{-1}$ Mpc$^{-1}$, $\Omega_m$ = 0.286, and $\Omega_{\Lambda}$ = 0.714, and we use Ned Wright's Javascript Cosmology Calculator website \citep{Wright06}. 

\begin{figure*}[htp]

\centering
	\includegraphics[angle=0,trim={1.5cm 1.5cm 11cm 0.8cm},clip=true,width=1\linewidth,keepaspectratio]{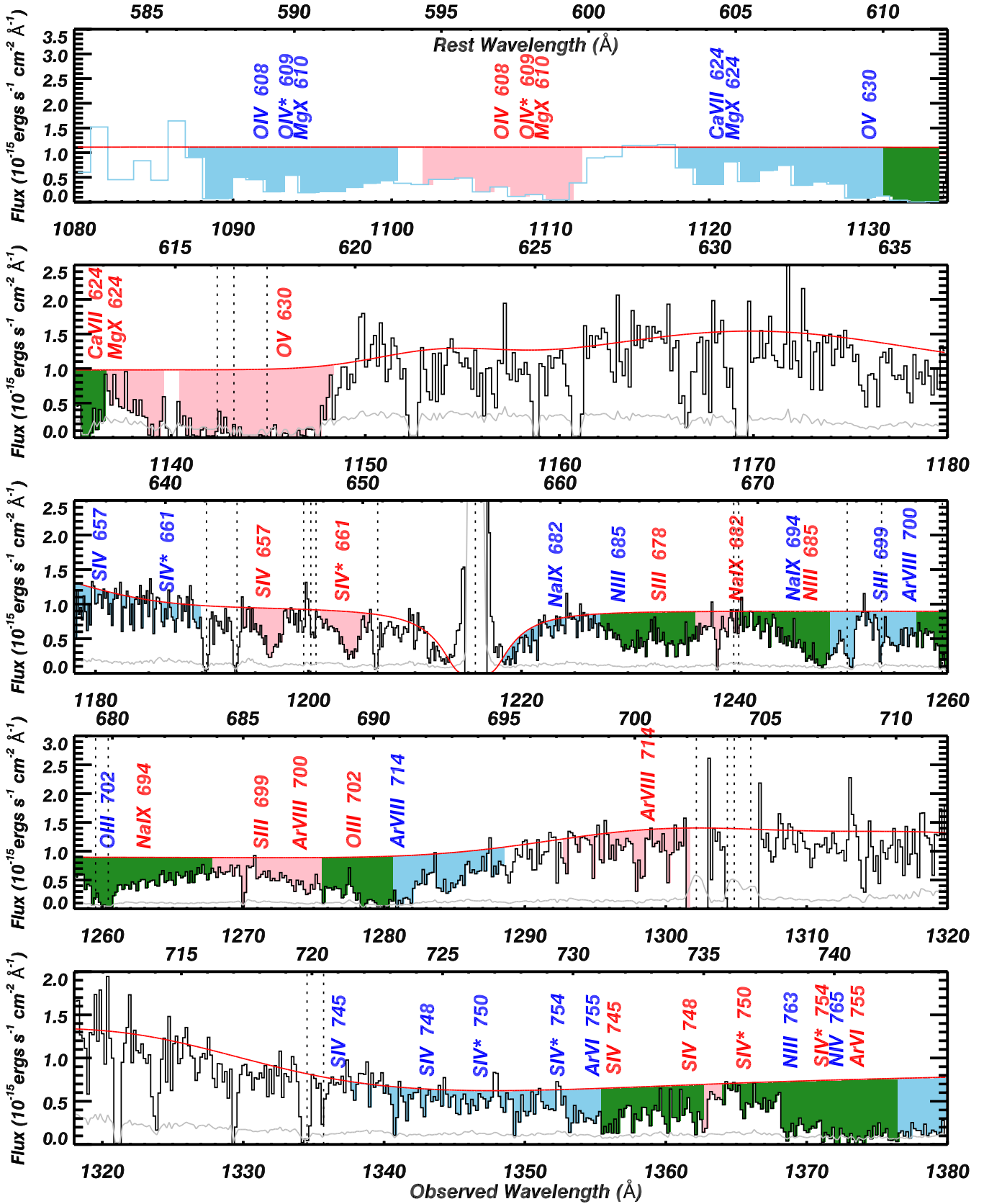}

\caption{HST/COS dereddened spectrum of SDSS J0755+2306 (z = 0.854). The black histogram shows the data from the 2017 epoch. The unabsorbed emission model and the flux error are shown as the red and gray solid lines, respectively. We shade the significant ionic absorption troughs for the two outflow \comps, S1 and S2, of the 2017 epoch in red and blue, respectively. Blended regions of the two outflow systems are shaded green. Strong Galactic interstellar medium (ISM) lines (e.g., \cii\ \ly 1334.53 and \cii*\ \ly 1335.71) and geocoronal lines (e.g., \hi\ at 1215.67\angstrom, \oi\ at 1302.17\angstrom, and \oi*\ at 1304.86\angstrom, 1306.03\angstrom) are marked with black dotted lines. The Galactic damped \Lya\ (at 1215.67\angstrom\ rest frame) is modeled by a Voigt profile with log(\Nh) = 20.4 cm$^{-2}$. The 2010 data are the blue histograms in the first panel. This covers an extra wavelength range from 1080 -- 1135\angstrom. The 2010 data are consistent with the 2017 data in the overlapping regions.
}

\label{fig:spec1}
\end{figure*}

\renewcommand{\thefigure}{\arabic{figure} (Continued)}
\addtocounter{figure}{-1}

\begin{figure*}[htp]

\centering
	\includegraphics[angle=0,trim={1.5cm 1.6cm 11cm 0.0cm},clip=true,width=1\linewidth,keepaspectratio]{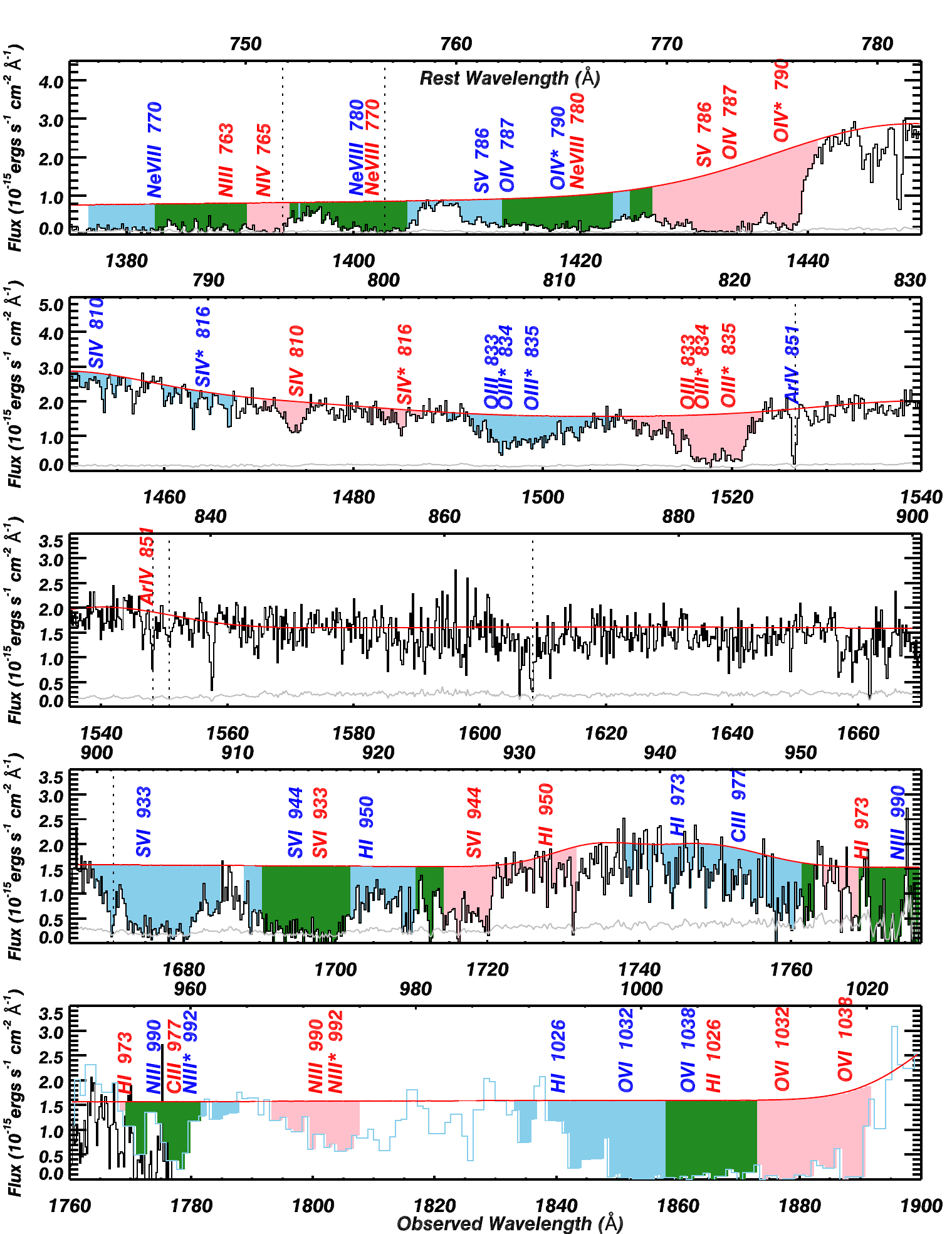}

\caption{The 2010 data are the blue histograms in the last panel. This covers an extra wavelength range from 1778 -- 1900\angstrom. The 2010 data are consistent with the 2017 data in the overlapping regions.
}
\end{figure*}

\renewcommand{\thefigure}{\arabic{figure}}

\begin{deluxetable}{c c c c c }[htp]
\tablewidth{0.45\textwidth}
\setlength{\tabcolsep}{0.02in}
\tablecaption{HST/COS Observations for SDSS J0755+2306}
\tablehead{
 \colhead{Epoch}	& \colhead{Date} & \colhead{Exp.$^{1}$} 	& \colhead{Grating} & \colhead{$\lambda_{c}^{2}$}
}

\startdata
1 	& 2017 Sep 18	&1220		&G130M	&	1291			\\
2 	& 2017 Sep 18	&2330		&G130M	&	1327		\\
3 	& 2017 Sep 19	&2330		&G160M	&	1577			\\
4 	& 2017 Sep 19	&2330		&G160M	&	1600			\\
5 	& 2010 Dec 20	&900		&G140L	&	1280			\\

\vspace{-2.2mm}
\enddata

\tablecomments{\\
$^{1}$: The exposure time of each observation in seconds.\\
$^{2}$: The central wavelength of each grating in\angstrom.}
\label{table: epochs}
\end{deluxetable}

\section{Observations and Data Reduction}
\label{sec:data}
SDSS J0755+2306 (J2000: R.A. = 07:55:14.58, decl. = +23:06:07.13, z = 0.854) was observed by HST/COS \citep{Green12} using gratings G130M and G160M in September of 2017 as part of our HST/COS program GO-14777 (PI: Arav). This object was observed previously in December of 2010 using the HST/COS G140L grating in the program GO-12289 (PI: J. Howk). The details of these observations are shown in table \ref{table: epochs}. We reduce and process the data and errors in the same way as described in \cite{Miller18}. We corrected for Galactic extinction with E(B-V) = 0.045 \citep{Schlafly12}. For the 2017 observations, we combined the two observations for each grating to increase the signal to noise. We show the full, dereddened spectrum in figure \ref{fig:spec1}. For regions outside the wavelength range of the 2017 epoch data, we show the 2010 epoch data.

Two outflow systems are identified: S1 has a velocity centroid ($v_c$) at --5520 km s$^{-1}$ (based on its \siv\ \ly 809.66 trough) and S2 at --9660 km s$^{-1}$ (based on its \svi\ \ly 933.38 trough). In figure \ref{fig:spec1}, absorption troughs associated with S1 and S2 are shaded in red and blue, respectively. Blended regions of the two outflow systems are shaded green. The unabsorbed emission model is comprised of a power law continuum and strong emission lines fitted with Gaussian profiles \citep{Chamberlain15b, Miller18, Xu18a}. The Galactic damped \lya\ absorption is modeled with a Voigt profile \citep[log(\Nh) = 20.4$^{+0.15}_{-0.15}$ cm$^{-2}$, e.g.,][]{Prochaska05}. The final, adopted emission model is shown as the solid red line in figure \ref{fig:spec1}.

\begin{deluxetable}{c c l c }[htb!]
\tablewidth{0.5\textwidth}
\tabletypesize{\small}
\setlength{\tabcolsep}{0.02in}
\tablecaption{Ionic Column Densities for Outflows in SDSS J0755+2306}
\tablehead{
	 \colhead{Ion}		& \colhead{ $\lambda$$^{(1)}$}		& \colhead{ N$_{ion,mea}$$^{(2)}$} 		 & \colhead{ $\frac{N_{ion,mea}}{N_{ion,model}}$$^{(3)}$  }  			
\\
\\ [-2mm]
	 \colhead{} 		& \colhead{(\AA)}		& \colhead{log(cm$^{-2}$)}	 		& \colhead{}		
}

\startdata

\multicolumn{4}{l}{\textbf{Outflow S1, $v$ = [-7000, -5200]$^{(4)}$}}\\ 
\hline	
			\hi		&949.74			&\color{red}$<$15.96			&$<$0.85		\\
			\niii		&685.52			&\color{blu}$>$15.55			&$>$1.12		\\
			\oiii		&832.93			&\color{blu}$>$16.02			&$>$1.20		\\
			\ov	 	&630.80			&\color{blu}$>$15.83			&$>$0.12			\\
			\ovi	 	&1037.62		&\color{blu}$>$16.30			&$>$0.24		\\
			\neviii		&780.32			&\color{blu}$>$16.49			&$>$1.00		\\
			\naix	 	&682.72			&\color{red}$<$15.71			&$<$31.6		\\
			\mgx	 	&624.94			&\color{blu}$>$16.69			&--$^{5}$		\\
			\siv+\siv*	&809.66, 815.94		&15.32$^{+0.12}_{-0.16}$		&0.85		\\
			\svi	 	&944.52			&\color{blu}$>$15.56			&$>$1.12		\\
			\ariv		&850.60			&\color{red}$<$14.78			&$<$1.51		\\
			\arviii		&713.80			&\color{red}$<$15.14			&$<$3.80		\\
			\cavii		&624.38			&\color{blu}$>$15.40			&--$^{5}$		\\
				 		
\hline

\multicolumn{4}{l}{\textbf{Outflow S2, $v$ = [-11200, 8000]$^{(4)}$}}\\ 
\hline
			\hi		&972.54			&\color{red}$<$16.15			&$<$1.02		\\
			\hi		&1025.72		&\color{blu}$>$15.68			&$>$0.35		\\
			\niii		&685.52			&\color{blu}$>$15.63			&$>$1.23		\\
			\oiii		&832.93			&\color{blu}$>$15.76			&$>$1.12		\\
			\ov	 	&630.80			&\color{blu}$>$15.98			&$>$0.06			\\
			\ovi	 	&1037.62		&\color{blu}$>$16.49			&$>$0.05		\\
			\neviii		&770.41			&\color{blu}$>$16.47			&$>$0.31		\\
			\naix	 	&682.72			&\color{red}$<$15.40			&$<$7.24		\\
			\mgx	 	&624.94			&\color{blu}$>$16.40			&--$^{5}$		\\
			\siv+\siv*	&809.66, 815.94		&\color{red}$<$15.37			&$<$1.35		\\
			\svi	 	&933.38			&\color{blu}$>$15.59			&$>$1.12		\\
			\ariv		&850.60			&\color{red}$<$14.68			&$<$1.78		\\
			\arviii		&700.24			&\color{blu}$>$15.39			&$>$0.41		\\
			\cavii		&624.38			&\color{blu}$>$15.70			&--$^{5}$		\\
\vspace{-2.2mm}

\enddata

\tablecomments{
\\
$^{1}$ The rest wavelength of the measured transitions for each ion. For ions which are a doublet or multiplet, we show all the uncontaminated transitions. \\
$^{2}$ The measured \Nion. Lower limits are shown in {\color{blu}blue} while upper limits are shown in {\color{red}red}. \siv+\siv*\ is for the sum of the resonance and excited transitions for \siv.\\
$^{3}$ The ratio of the measured \Nion\ to the model predicted \Nion.\\
$^{4}$ The N$_{ion}$ integration range in km s$^{-1}$.\\
$^{5}$ For the transitions of \mgx\ \ly 624.94 and \cavii\ \ly 624.38, their absorption troughs are too close to be disentangled. We report the \Nion\ values for \mgx\ or \cavii\ assuming that the whole blended trough is from \mgx\ or \cavii, respectively (see section \ref{sec:Nion}). In the photoionization models, we investigated several possible scenarios (section \ref{sec:model}).
}

\label{tb:IonSystems2}
\end{deluxetable}

\section{ Spectral Analysis}
\label{sec:spec}

\subsection{Column Density Determinations}
\label{sec:Nion}
The ionic column densities (\Nion) measured from the spectra represent the ionization structure of the observed outflow material. Like in all 11 outflows in the other objects (see table 1 of Paper I), we observe in S1 and S2 strong absorption troughs from very high-ionization species, including \arviii, \neviii, \naix, and \mgx. Similar to 10 of the outflows in the other objects (the exception is S4 of SDSS J1042+1646, see Paper IV), we observe in S1 and S2 absorption troughs from triply ionized species, e.g., \niv, \oiv, and \siv. We also observe absorption troughs from doubly ionized species in S1 and S2, including \ciii, \niii, and \oiii. The only other outflow analyzed in our EUV500 program that shows such troughs is the --3150 km s$^{-1}$ outflow system in Paper V. Overall, we observe troughs in quasar SDSS J0755+2306 from ions with a larger spread of ionization potentials (48 eV $\sim$ 367 eV) than in most of the other analyzed outflows in our EUV500 program. The atomic data for these transitions are shown in table 3 of Paper II.

Following the methodology in section 3 of Paper II, we analyze the data and measure \Nion\ as follows. Most measured \Nion\ use the apparent optical depth (AOD) method. Visual inspection of the troughs between epochs show no significant variability. Therefore, when possible, we use the \Nion\ measurements from the 2017 epoch data since it has higher signal-to-noise and spectral resolution. Most of the measured troughs are treated as lower limits since their levels of non-black saturation are unknown without available partial covering (PC) solutions \citep{Borguet12b}. For absorption trough regions with a maximum optical depth, $\tau_{max}$ $<$ 0.05, we consider their AOD \Nion\ as upper limits. In section \ref{sec:distance}, we show that we can obtain \Nion\ measurements for \siv\ and \siv* for S1. We show the measured \Nion\ in the third column of table \ref{tb:IonSystems2} and the corresponding ion and wavelength in the first two columns. All troughs in figure \ref{fig:spec1} that are not listed in table \ref{tb:IonSystems2} are severely blended, yielding unreliable \Nion\ measurements or limits.

For the transitions of \mgx\ \ly 624.94 and \cavii\ \ly 624.38, their absorption troughs are too close to be disentangled (S1 around 1135\angstrom\ and S2 around 1120\angstrom, observed-frame). In table \ref{tb:IonSystems2}, we report the \Nion\ values for \mgx\ or \cavii\ assuming that the whole blended trough is from \mgx\ or \cavii, respectively. When determining the photoionization solutions, we investigate several possible scenarios for the blending between \mgx\ and \cavii\ (see section \ref{sec:model}).

For the \Nion\ of \hi\ in S2, the ionic transition of \hi\ \ly 972.54 does not show consistently deep absorption trough features near 1745\angstrom\ observed-frame. Therefore, we measure the AOD \Nion\ from the trough of \hi\ \ly 972.54 and treat it as an upper limit for \hi. The ionic transition of \hi\ \ly 1025.72 exhibits deep absorption near 1840 -- 1845\angstrom\ observed-frame, while the right wing is blended with \ovi\ \ly 1031.91. We assume the trough from \hi\ \ly 1025.72 is symmetric and double its blue half AOD value for the lower limit \Nion\ of \hi. Therefore, the \Nion\ of \hi\ for outflow S1 is constrained to the range of 15.68 to 16.15 [in units of log(cm$^{-2}$)].

\begin{figure}[htp]

\centering
	\includegraphics[angle=0,trim={0cm 1.80cm 0.3cm 1.5cm},clip=true,width=1\linewidth,keepaspectratio]{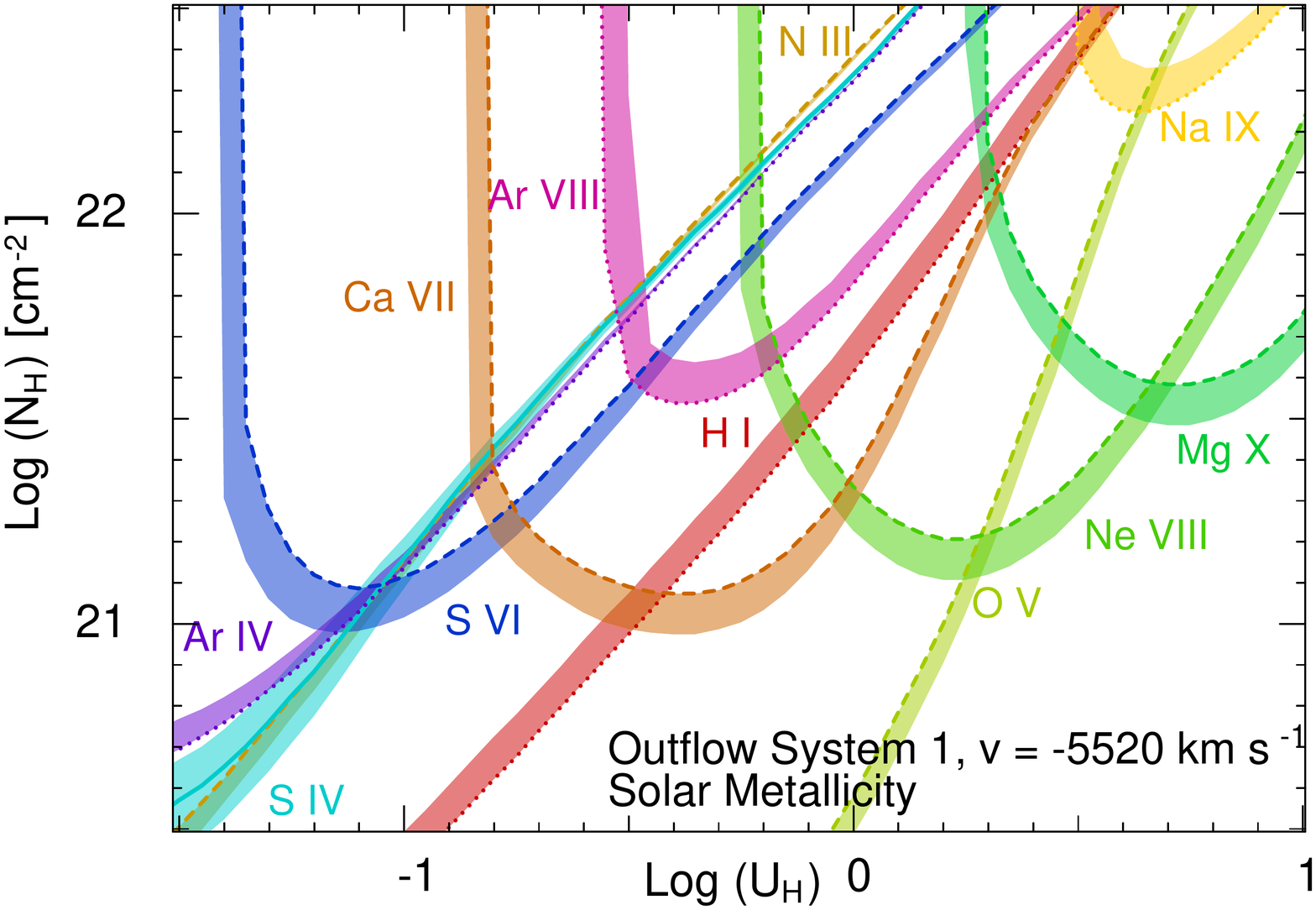} 
	\includegraphics[angle=0,trim={0cm 0cm 0.3cm 1.5cm},clip=false,width=1\linewidth,keepaspectratio]{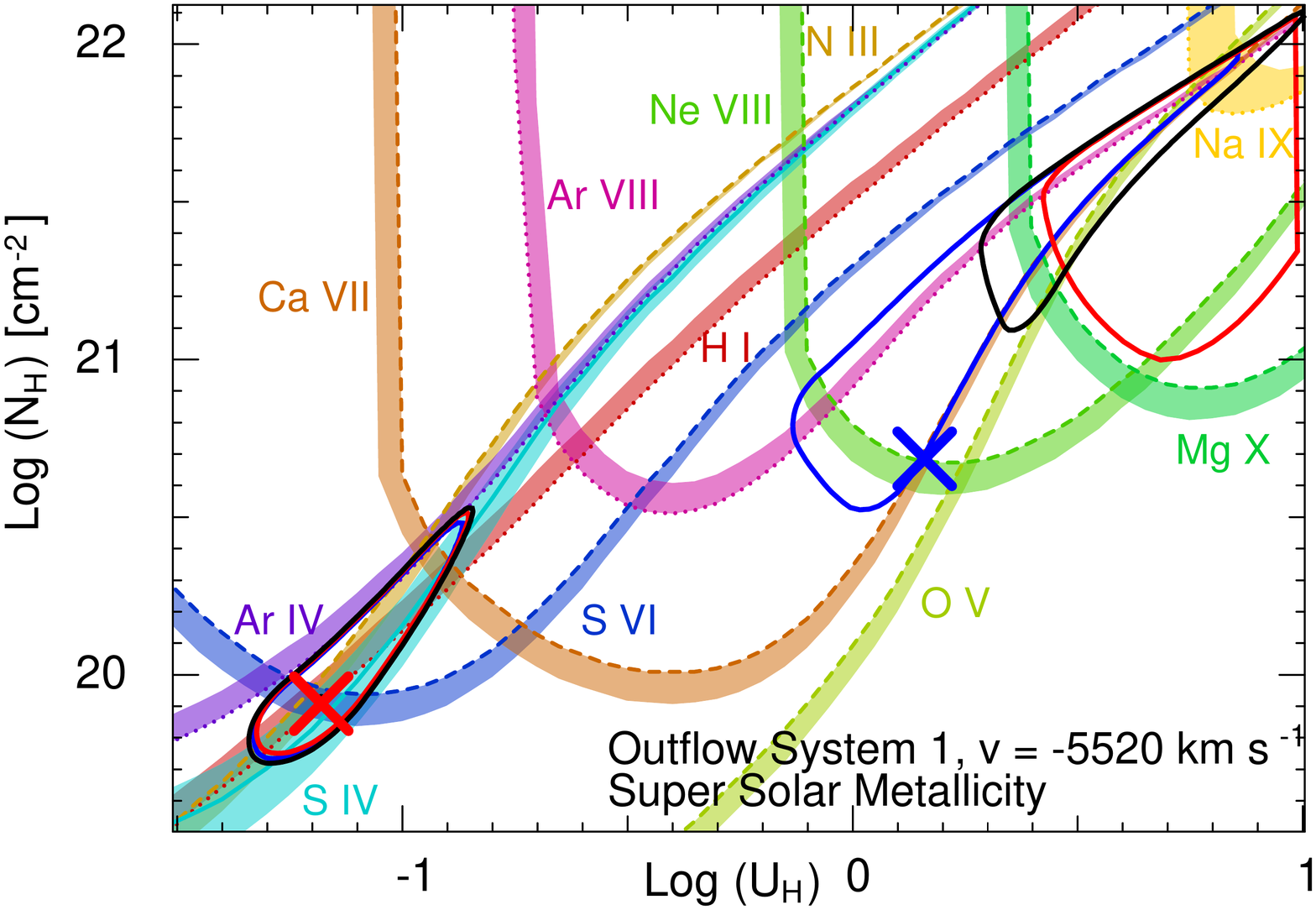} 
	\includegraphics[angle=0,trim={0cm 0cm 0.3cm 1.70cm},clip=true,width=1\linewidth,keepaspectratio]{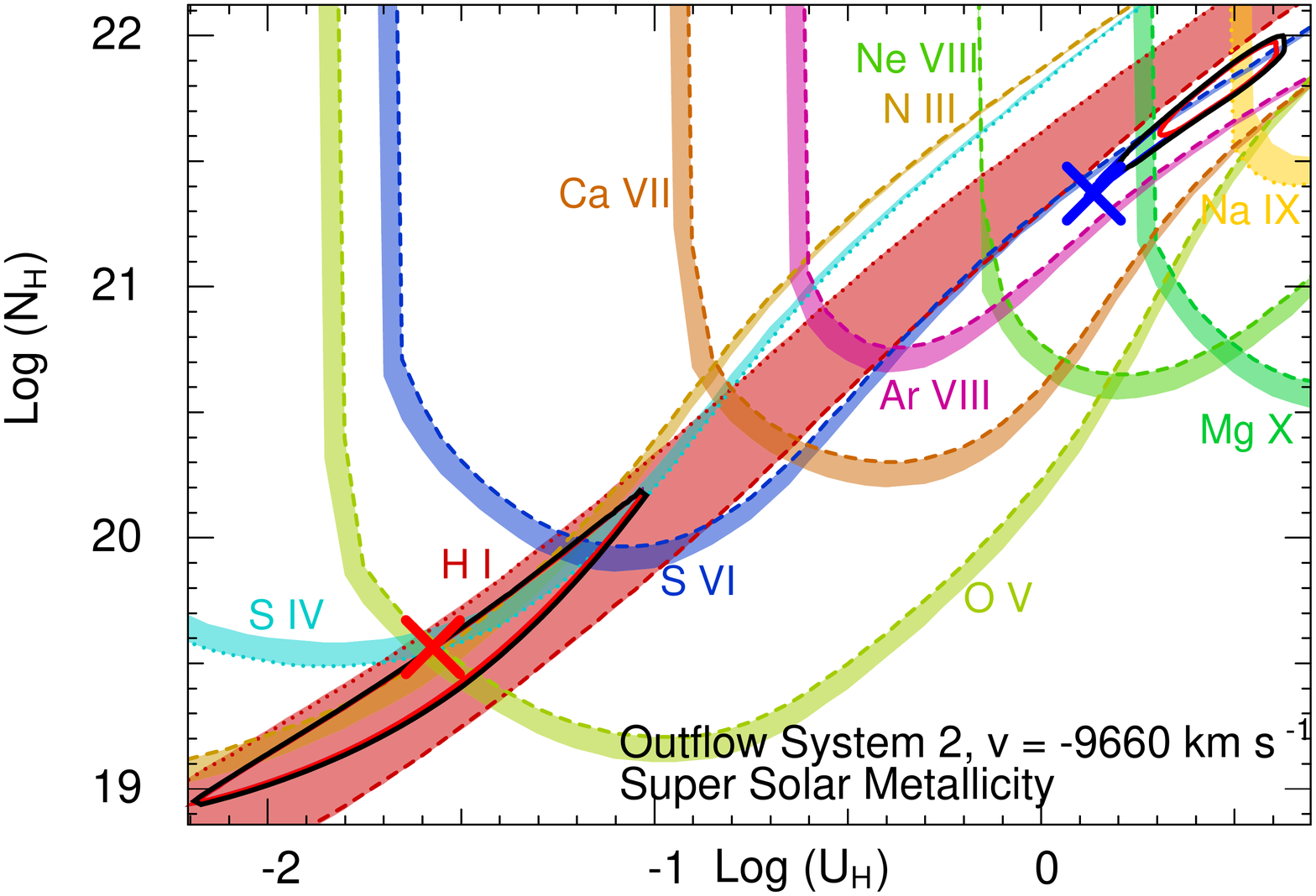} 

\caption{The best fitting photoionization solutions for outflows S1 and S2. \textbf{Top:} Comparison of the Cloudy modeled N$_{ion}$ to the measured N$_{ion}$  in S1 assuming solar metallicity. Each colored contour represents the region where the (\Nh, \Uh) model produces consistent N$_{ion}$ within the errors with the observed values. Solid lines represent N$_{ion}$ measurements while dotted and dashed lines represent upper and lower \Nion\ limits, respectively. Any solution that matches the upper limit N$_{ion}$ of \hi\ underpredicts N$_{ion}$ of \niii, \siv, and \svi\ by up to a factor of 5.  \textbf{Middle and Bottom:} Under super-solar metallicity, the N$_{ion}$ from S1 and S2 match with two-phase photoionization solutions (see section \ref{sec:model}). The very high- and high-ionization phase solutions are the blue and red ``$\times$" along with their 1$\sigma$ error contours (the black ellipses), respectively. The black, blue, and red ellipses are accounting for the blending of troughs from \mgx\ \ly 624.94 and \cavii\ \ly 624.38 (see section \ref{sec:model}). The other \Nion\ lower and upper limits which are not shown here are consistent with the solutions and omitted for clarity's sake. }

\label{fig:comp1}
\end{figure}

\subsection {Photoionization Analysis}
\label{sec:model}
We assume the spectral energy distribution HE0238 SED \citep{Arav13}. This SED is physically plausible since it is based on observations of quasar HE 0238--1904 in the EUV500 band \citep{Arav13}. Two main parameters govern the photoionization structure of each outflow: the total hydrogen column density (\Nh) and the ionization parameter (\Uh):

\begin{equation}
\label{Eq:ionPoten}
\Uh \equiv\ \frac{\Qh}{4\pi R^2 \nh c}
\end{equation}
where $R$ is the distance from the central source to the absorber, $\nh$\ is the hydrogen number density, $c$ is the speed of light, and $\Qh$ = 3.1 $\times$ 10$^{56}$ s$^{-1}$ is the emission rate of hydrogen-ionizing photons (obtained by integrating the HE0238 SED for energies above 1 Ryd). The corresponding bolometric luminosity is $\sim$ 4.4 $\times$ 10$^{46}$~erg~s$^{-1}$.


We start with assuming solar metallicity and compare the measured N$_{ion}$ (table \ref{tb:IonSystems2}) to the model predicted N$_{ion}$ from the spectra synthesis code Cloudy [version c17.00, \cite{Ferland17}] (top panel of figure \ref{fig:comp1}). The colored contours for individual ions show where the measured N$_{ion}$ are consistent ($\leq$ 1$\sigma$) with the modelled N$_{ion}$ from Cloudy \citep{Borguet12a}. The colored contours with solid lines are N$_{ion}$ measurements, and dotted or dashed lines are N$_{ion}$ upper or lower limits, respectively. It is evident that there is no viable solution for solar metallicity. Any solution that matches the upper limit N$_{ion}$ of \hi\ will simultaneously underpredict N$_{ion}$ of \niii, \siv, and \svi\ by up to a factor of 5. 

One possible solution is to invoke a super-solar metallicity. There were outflow systems which have been found to have super-solar metallicity \citep[e.g.,][]{Gabel06,Arav07,Arav19}. In the middle and bottom panel of figure \ref{fig:comp1}, we present the photoionization solutions assuming the HE0238 SED and the super-solar metallicity described in Paper V (Z = 4.68 $Z_{\odot}$). As in most of the other EUV500 outflows in our HST program GO-14777 (see table 1 of Paper I, except S4 in SDSS J1042+1646), we invoke a two-phase photoionization solution for both S1 and S2 \citep{Arav13}. The very high-ionization phase (VHP) and high-ionization phase (HP) solutions are the blue and red ``$\times$" along with their 1$\sigma$ error contours (the black ellipses), respectively. The ratios of the measured \Nion\ to the model predicted \Nion\ are given in the fourth column of table \ref{tb:IonSystems2}. When N$_{ion,mea}$ is a lower limit, we expect to have N$_{ion,mea}$/N$_{ion,model}$ $<$ 1 and vice versa.

Due to the blending absorption troughs from \mgx\ \ly 624.94 and \cavii\ \ly 624.38 (S1 around 1135\angstrom\ and S2 around 1120\angstrom\ observed-frame, see figure \ref{fig:spec1}), we present the photoionization solutions considering three different blending scenarios (figure \ref{fig:comp1}). 1) half the trough's optical depth is from the \caviii\ ionic transition and the other half is from the \mgx\ ionic transition black ellipses), 2) the trough is comprised of only the ionic transition of \cavii\ \ly 624.38 (blue ellipses), and 3) the trough is contributed from only the ionic transition of \mgx\ \ly 624.94 (red ellipses). For both S1 and S2, the blue ``$\times$" denotes the photoionization solution with the least \Nh\ for the very high-ionization phase. 


\begin{figure}[htp]
	
	\centering
	\includegraphics[angle=0,trim={0cm 0cm 0.0cm 7cm},clip=true,width=1\linewidth,keepaspectratio]{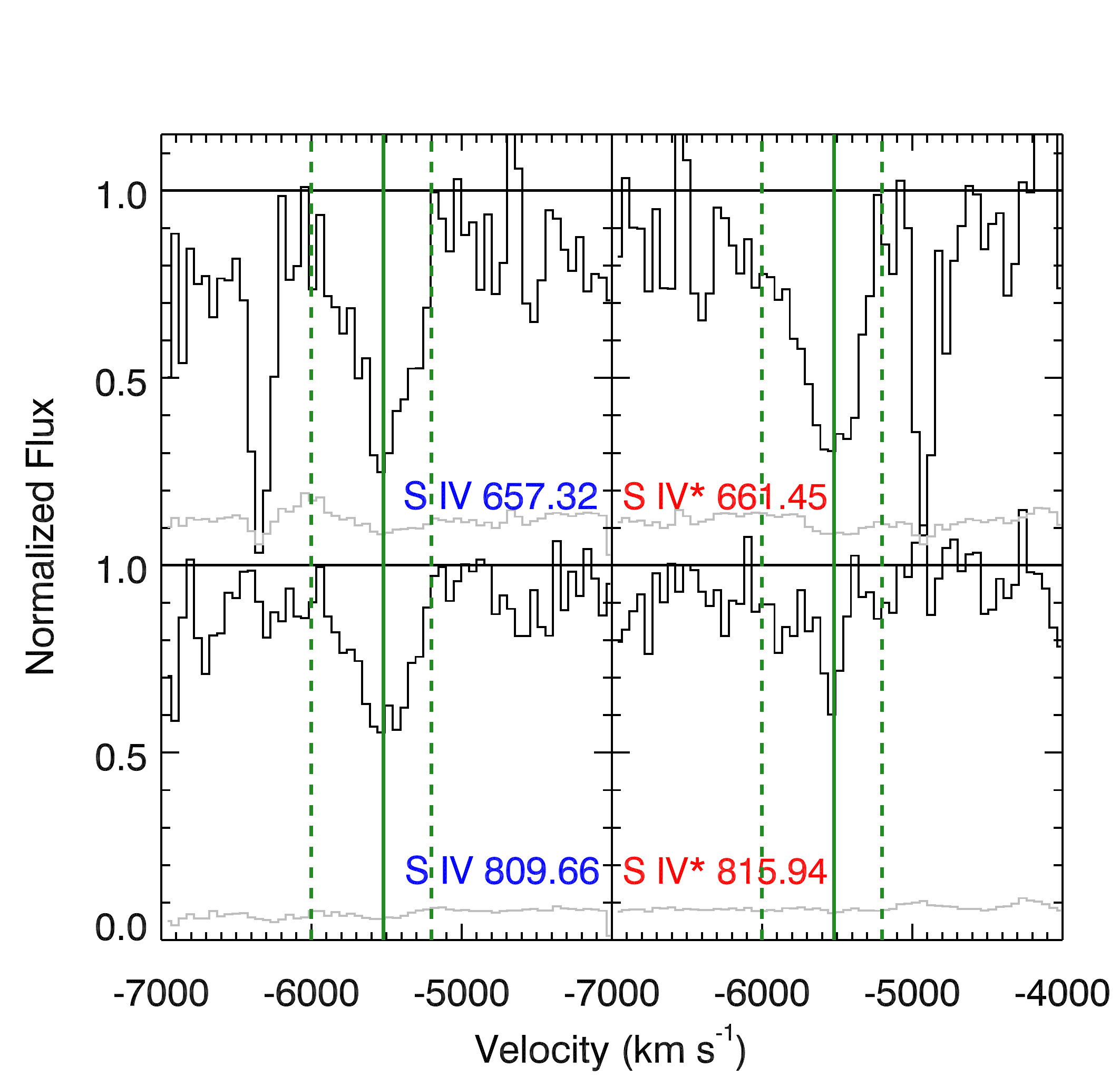} 
	\caption{Comparison between two pairs of \siv\ and \siv*\ troughs for S1. The data are shown as the black histogram. The vertical, green solid line shows the velocity centroid of S1, while the \Nion\ integration range are shown as the dashed green lines (see section \ref{sec:distance}).
	}
	\label{vcut:comp1}
\end{figure}

\section{Electron Number Density and Distances }
\label{sec:distance}
By assuming the outflow is governed by photoionization, we can solve for $R$ from equation (\ref{Eq:ionPoten}). The only other unknown parameter is \nh, and in a highly ionized plasma, \nh\ $\approx$ 0.8\ne. Here we use the density sensitive \Nion\ ratio from \siv*/\siv\ (for S1) and \oiii*/\oiii\ (for S2) to constrain \ne.


\begin{deluxetable}{l c r c}[]
\tablewidth{0.45\textwidth}
\setlength{\tabcolsep}{0.02in}
\tablecaption{Atomic Data for \siv\ and \siv*\ Transitions}
\tablehead{
 \colhead{Ion}	& \colhead{$\lambda$$^{(1)}$} 	& \colhead{\enspace\enspace E$_{low}$$^{(2)}$}	& \colhead{f$^{(3)}$} 	\\ 
 \colhead{}		& \colhead{(\AA)} 		& \colhead{\enspace\enspace(cm$^{-1}$)}		& \colhead{} 		
}
\startdata
\siv			& 657.319	&0.00		&1.130		\\	
\siv*			& 661.396	&951.4		&1.130		\\
\siv			& 744.904	&0.00		&0.249		\\
\siv			& 748.393	&0.00		&0.459		\\
\siv*			& 750.221	&951.4		&0.597		\\
\siv*			& 753.760	&951.4		&0.131		\\
\siv			& 809.656	&0.00		&0.118		\\
\siv*			& 815.941	&951.4		&0.085		

\enddata
\tablecomments{\\
$^{(1)}$ Rest wavelength of \siv\ and \siv*\ transitions. \\
$^{(2)}$ Lower-level energy of these transitions from the National Institute of Standards and Technology (NIST) database \citep{NIST18}.\\
$^{(3)}$ Oscillator strengths from the NIST database.\\
}
\label{tab:sivad}
\end{deluxetable}

\subsection{Determination of \ne\ for S1 from \siv*/\siv}
For S1, we observe absorption at the expected wavelength locations of the \siv\ lines listed in table \ref{tab:sivad}. However, the 744.90\angstrom, 748.39\angstrom, 750.22\angstrom, and 753.76\angstrom\ troughs are severely blended with absorption troughs from S2 (see figure \ref{fig:spec1}). Therefore, the \Nion\ from these \siv\ transitions can not be reliably determined. The 657.32\angstrom, 661.40\angstrom, 809.66\angstrom, and 815.94\angstrom\ troughs are not blended with other troughs from S2 or strong intervening systems (see figure \ref{vcut:comp1}). We show the comparison of these troughs in velocity space in figure \ref{vcut:comp1}.

The velocity centroids match well for these troughs as indicated by the green solid lines, while the \Nion\ integration ranges are the green dotted lines. The 815.94\angstrom\ trough clearly has less \Nion\ than the 809.66\angstrom\ trough, which is consistent with our derived N(\siv*)/N(\siv) ratio. For the AOD method, the expected optical depth ($\tau$) ratio of the 657.32\angstrom\ trough to the 809.66\angstrom\ trough is:
\begin{equation}\label{eq:1}
\begin{split}
\frac{\int \tau(v)_{657.32} dv}{\int \tau(v)_{809.66} dv} = \frac{N_{S\text{$\scriptstyle{IV}$}} \times f_{657.32} \times {657.32}}{N_{S\text{$\scriptstyle{IV}$}} \times f_{809.66} \times {809.66}} = 7.8
\end{split}
\end{equation}

where N(\siv) is the column density of \siv\ and $f_{657.32}$/$f_{809.66}$ $\simeq$ 9.6 is the oscillator strength ratio between the two transitions. By assuming that $\tau(v)_{657.32}$/$\tau(v)_{809.66}$ equals a constant, the expected ratio in the AOD case, i.e., [$\tau(v)_{657.32}$/$\tau(v)_{809.66}$]$_\text{AOD}$, is 7.8. However, the observed ratio, i.e., [$\tau(v)_{657.32}$/$\tau(v)_{809.66}$]$_\text{obs}$, is around 3, which indicates that the 657.3\angstrom\ trough is non-black saturated. Similarly, for the excited states of \siv, we derived [$\tau(v)_{661.40}$/$\tau(v)_{815.94}$]$_\text{AOD}$  =  9.7 and [$\tau(v)_{661.40}$/$\tau(v)_{815.94}$]$_\text{obs}$ $\simeq$ 3. Therefore, the 661.40\angstrom\ trough is also non-black saturated. Thus, we use the PC method to obtain the \Nion\ for the \siv\ resonance state (E$_\text{low}$ = 0 cm$^{-1}$) from the 657.32\angstrom\ and 809.66\angstrom\ troughs and the excited state from the 661.40\angstrom\ and 815.94\angstrom\ troughs. The resulting ratio of the \siv* column density to the \siv\ column density, i.e., N(\siv*)/N(\siv), is 0.54$^{+0.20}_{-0.17}$. 

\begin{figure}[htp]
	\centering
	\includegraphics[angle=00,trim={0.5cm 0.0cm 0cm 12.5cm},clip=true,width=1\linewidth,keepaspectratio]{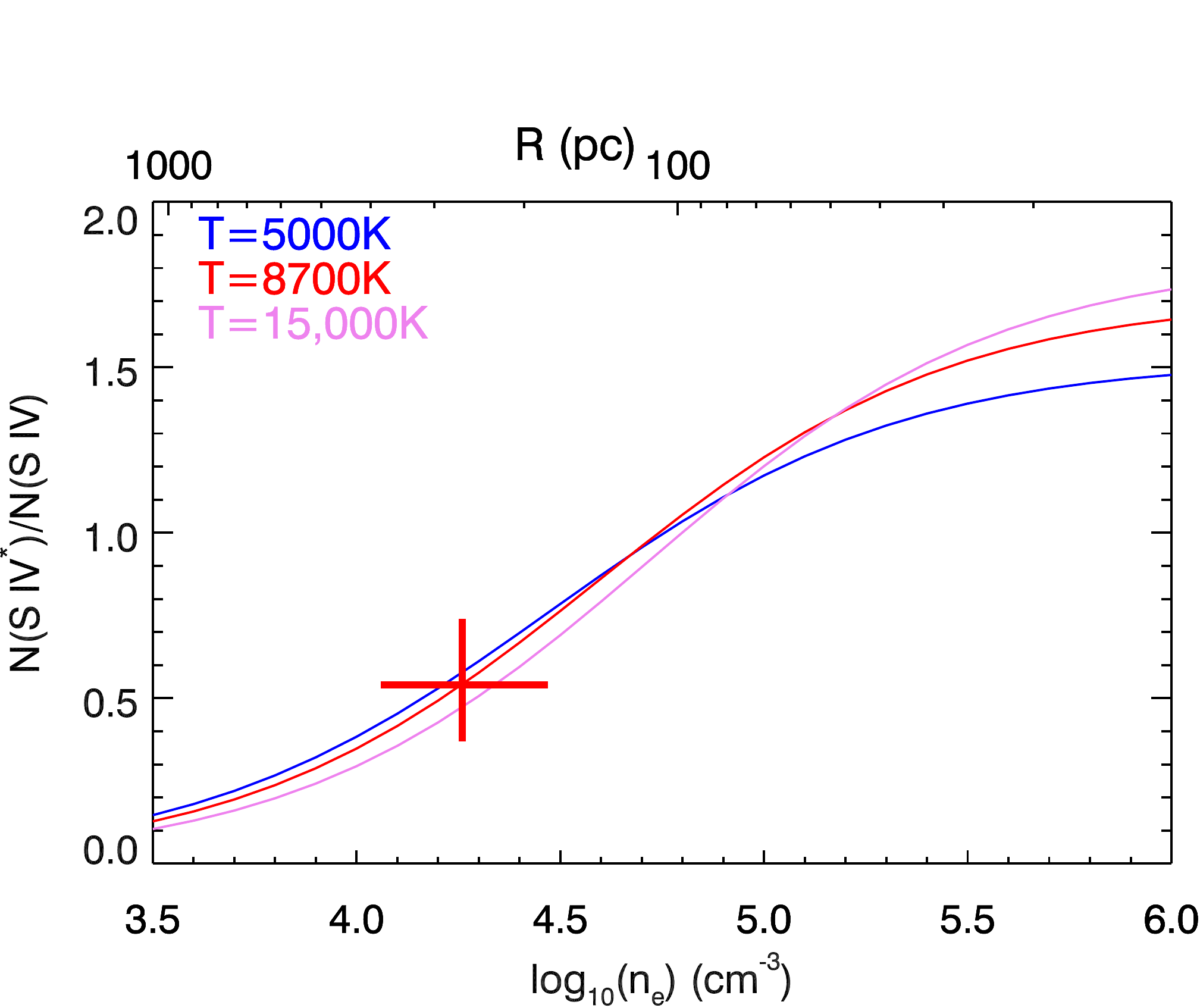}
	\caption{Column density ratio of \siv*\ to \siv\ vs. electron number density, \ne, for outflow S1. The red cross marks the ratio with the uncertainties derived in section \ref{sec:distance}. The colored curves are the predictions from the CHIANTI database \citep[version 7.1.3,][]{Landi13} assuming different temperature. The mean temperature of the \siv\ gas based on the photoionization solution for the HP of S1 is 8700 K (section \ref{sec:model}).  }
	
	\label{fig:SIV}
\end{figure}

In figure \ref{fig:SIV}, we compare this \siv\ column density ratio to those predicted by the CHIANTI database \citep[version 7.1.3,][]{Landi13}. The mean temperature for \siv\ is 8700 K, which is based on the photoionization solution for the HP of S1 (section \ref{sec:model}). The red curve is the model predictions from CHIANTI while the red cross is the derived N(\siv*)/N(\siv) ratio with its uncertainties. We find log(\ne) = 4.26$^{+0.21}_{-0.20}$ (hereafter, \ne\ is in units of  log(cm$^{-3}$)). 


For outflow S1, we also observe absorption troughs from other density sensitive transitions, e.g., \oiv\ \ly 787.71 and \oiv*\ \ly 790.20 and \oiii+\oiii*\ near 833\angstrom\ in the observed frame (see figure \ref{fig:spec1}). Unfortunately, their absorption troughs are either saturated or too blended to provide useful \ne\ constraints. However, their absorption troughs are consistent with our best fitting photoionization model. Therefore, by adopting the best-fit \Uh\ and \siv-determined \ne\ into equation (\ref{Eq:ionPoten}), we obtain $R$ = 270$^{+100}_{-90}$ pc. 

\subsection{Determination of \ne\ for S2 from \oiii*/\oiii}
\label{subsec:ne_S2}
For outflow S2, the stronger \siv\ and \siv*\ transitions at 657.32\angstrom\ and 661.40\angstrom\ do not show distinctive troughs. However, we detect deep absorption features at the expected wavelength location of the \oiii+\oiii*\ multiplet (\oiii\ \ly 832.93, and \oiii*\ \ly\ly 833.75, 835.29). To determine \ne, we adopt the same analysis method from Paper II.

\begin{figure}[htp]
	\centering
	\includegraphics[angle=00,trim={0.5cm 7.0cm 0cm 3.5cm},clip=true,width=1\linewidth,keepaspectratio]{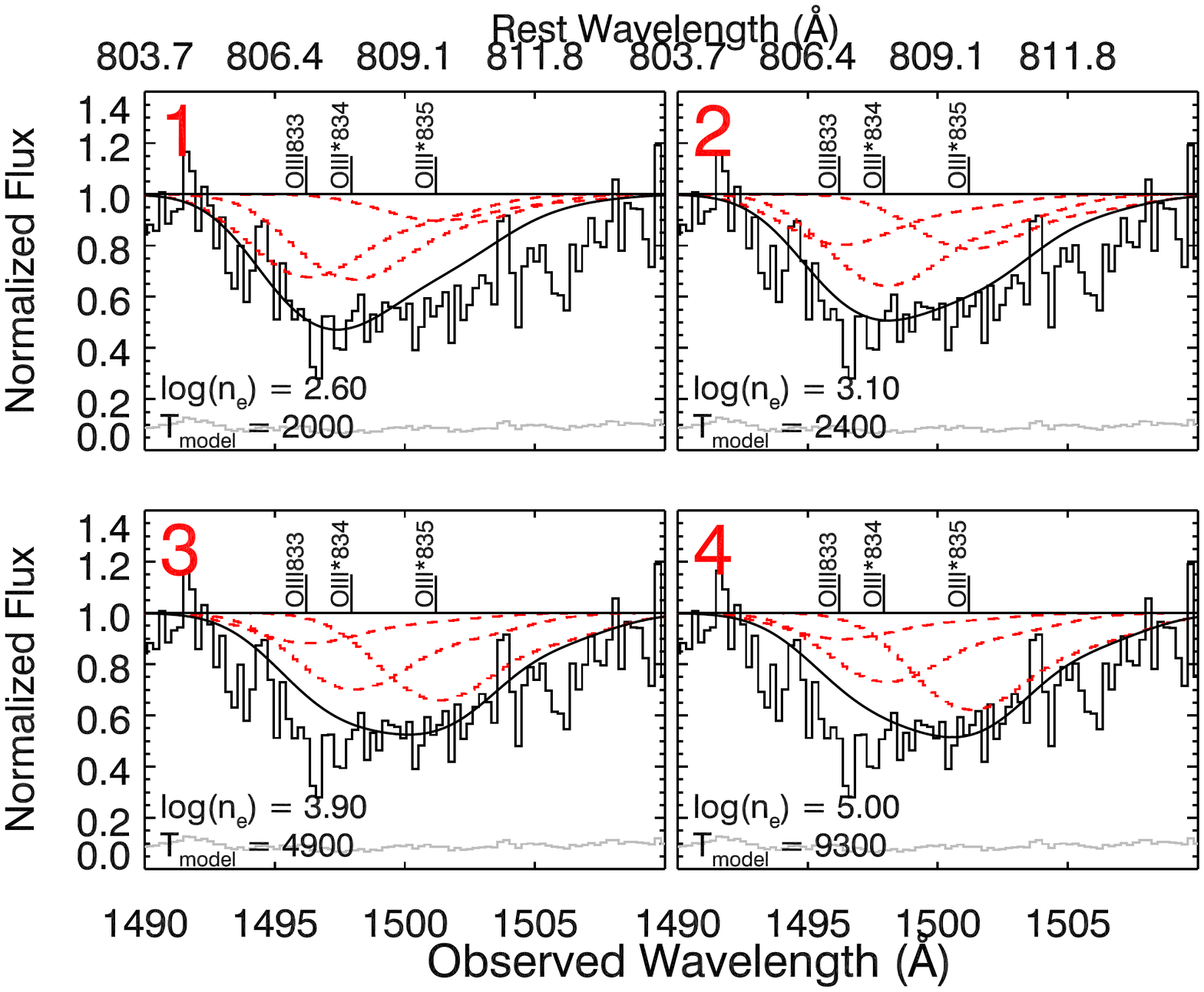}
	\caption{Fits to the \oiii+\oiii*\ multiplet region for outflow S2. To get the best fit, we vary \ne\ (in units of cm$^{-3}$) and probe the photoionization solution inside the 1$\sigma$ error contour of the HP in S2. The \ne\ and the corresponding temperature predicted from Cloudy are shown at the bottom-left corner of each panel. The black and gray solid histograms are the normalized flux and errors from the HST/COS observations in 2017. For each subplot, the red dashed lines represent the models of the \oiii+\oiii*\ multiplet for a particular log(\ne), while the solid black lines are the summation of all models in this region. See section \ref{subsec:ne_S2} for a detailed discussion.}
	
	\label{fig:OIII}
\end{figure}

To fit the observed absorption features, we start with the photoionization solution inside the 1$\sigma$ contour of the HP for S2 (the contour surrounding the red $\times$ in the bottom panel of figure \ref{fig:comp1}). We vary log(\ne) from 2 to 8 and overlay the model predicted \oiii+\oiii*\ troughs to the 1490 -- 1510\angstrom\ observed-frame region (see figure \ref{fig:OIII}). We then do a $\chi^2$-minimization of the data and model for the \oiii+\oiii*\ region. The red dashed lines represent the modeled troughs of the \oiii+\oiii*\ multiplet for a particular \ne, while the solid black lines are the summation of all models in the region.

Since a single-Gaussian optical depth profile (e.g., equation (2) of Paper II) does not fit the \oiii+\oiii*\ region well, we adopt a two-Gaussian optical depth profile following \cite{Borguet12b}. The two Gaussians have the same velocity width ($\sigma$) of 350 km s$^{-1}$. The main Gaussian contains 65\% of the total \Nion\ and has a velocity centroid (v$_{c}$) of --9660 km s$^{-1}$, while the secondary Gaussian contains 35\% of the total \Nion\ with v$_{c}$ = --8860 km s$^{-1}$. The same two-Gaussian profile also fits well the lower-velocity wing of other outflow troughs in S2, e.g., from \arviii\ \ly 713.80.  

Adopting the two-Gaussian profile, the best-fitting log(\ne) = 3.1, where the corresponding models are shown in the panel 2 of figure \ref{fig:OIII}. The models with log(\ne) = 2.6 and log(\ne) = 3.9 deviate from the best-fitting model by 1$\sigma$ (see panel 1 and 3 of figure \ref{fig:OIII}), where they clearly underestimate the absorption troughs from 1500 -- 1510\angstrom\ and 1493 -- 1498\angstrom\ observed-frame, respectively. Overall, we get log(\ne) = 3.1$^{+0.8}_{-0.5}$.


By adopting the best-fitting \ne\ value and errors into equation (\ref{Eq:ionPoten}), we obtain $R$ = 1600$^{+2000}_{-1100}$ pc.

\section{Outflow Energetics}
\label{sec:energy}
By assuming each outflow is in the form of a thin shell, covering a solid angle of $4\pi\Omega$ around the source, moving with a radial velocity $v$ at a distance $R$ from the central source \citep[see Paper I and][]{Borguet12a}, the mass flow rate ($\dot{M}$) and kinetic luminosity ($\dot{E}_{k}$) of the outflow are given by:

\begin{equation}\label{eq:1}
\begin{split}
\dot{M}\simeq 4\pi \Omega R\Nh \mu m_p v, \ \ \ \ \ 
\dot{E}_{k}\simeq \frac{1}{2} \dot{M}v^2
\end{split}
\end{equation}

where \Nh\ is the total hydrogen column density, m$_{p}$ is the proton mass, and $\mu$ = 1.4 is the mean atomic mass per proton. 

Using $R$ with the \Uh\ and \Nh\ from the best-fitting photoionization solutions, we present the derived $\dot{M}$ and $\dot{E}_{k}$ values in table \ref{tab:res}, where we assume $\Omega$ = 0.2 (see section 6.4 of Paper II). 

Using SDSS data, we measure the full-width-half-maximum of the \mgii\ broad emission line and estimate the Eddington luminosity (L$_{edd}$) with the \mgii--based black hole mass equation in \cite{Bahk19}. This leads to L$_{edd}$ = 1.0 $\times$ 10$^{47}$ erg s$^{-1}$. Therefore, outflows S1 and S2 yield the ratio of kinetic luminosity to L$_{edd}$ of $>$ 0.2\% and 12 -- 250\%, respectively. The large range for S2 is due to the uncertainties of its \Nh\ and \ne, while the conservative lower limit of 12\% is assured. Outflow S2, with $\dot{E}_{k}$ greater than 5\% of L$_{edd}$, is a good candidate for producing strong AGN feedback \citep{Scannapieco04}.


\begin{deluxetable}{lcccc}
\tabletypesize{\footnotesize}
\tablecaption{Physical Properties of the Outflow Systems Seen in \\Quasar SDSS J0755+2306\label{tab:res}}
\tablewidth{0pt}
\tablehead{
\colhead{Outflow System} & \multicolumn{2}{c}{$-$5520~km~s$^{-1}$ (S1)} & \multicolumn{2}{c}{$-$9660~km~s$^{-1}$ (S2)}\\
\tableline
\\
\colhead{Ionization Phase} & \colhead{Very High} & \colhead{High} & \colhead{Very High} & \colhead{High}
}
\startdata
log$_{}$(\sub{N}{H})  	& $>$ 20.7 & 19.9$^{+0.6}_{-0.2}$ 	& 21.4 -- 22.0 	& 19.5$^{+0.7}_{-0.6}$\\
\ [cm$^{-2}$]\\
\tableline
log$_{}$(\sub{U}{H})  	& $>$ 0.1 & -1.2$^{+0.3}_{-0.2}$ 	& 0.1 -- 0.6	& -1.6$^{+0.6}_{-0.5}$\\
\ [dex]\\
\tableline
log(\sub{n}{e}) 	& \tablenotemark{a}$<$3.0 		& 4.3$^{+0.2}_{-0.2}$& \tablenotemark{a}$<$1.8 & 3.1$^{+0.8}_{-0.5}$  \\
\ [cm$^{-3}$]\\
\tableline
Distance 		& \multicolumn{2}{c}{270$^{+100}_{-90}$}	& \multicolumn{2}{c}{1600$^{+2000}_{-1100}$}  \\
\ [pc]\\
\tableline
$\dot{M}$ 		& \multicolumn{2}{c}{$>$ 20}	& \multicolumn{2}{c}{450 -- 8000}  \\
\ [$M_{\odot}$ yr$^{-1}$]\\
\tableline
log($\dot{E_{k}}$)\tablenotemark{b} & \multicolumn{2}{c}{$>$44.3}& \multicolumn{2}{c}{46.1 -- 47.4} \\
\ [erg s$^{-1}$]\\
\tableline
$\dot{E_{k}}/L_{edd}$ & \multicolumn{2}{c}{$>$0.002}	& \multicolumn{2}{c}{0.125 -- 2.5} \\
\ \\
\tableline
log(\ssub{f}{V})\tablenotemark{c} 	& \multicolumn{2}{c}{$<$-2.1}	& \multicolumn{2}{c}{$<$-3.3}  \\
\enddata
\tablecomments{Bolometric luminosity, \sub{L}{bol} = 4.4$\times$10$^{46}$ erg s$^{-1}$ assuming the HE0238 SED.\\
(a) Assuming that both ionization components are at the same distance.\\
(b) Assuming $\Omega$ = 0.2 and where $N_H$ is the sum of the two ionization phases.\\
(c) The volume filling factor of the outflow's high-ionization phase to the very high-ionization phase \citep[see table 1 in Paper I and ][]{Arav13}.
}
\end{deluxetable}

\section{Discussion}
\label{sec:discussion}

\begin{figure*}[htp]
\centering
    \includegraphics[angle=00,trim={1.5cm 0.5cm 11.5cm 11.5cm},clip=true,width=1\linewidth,keepaspectratio]{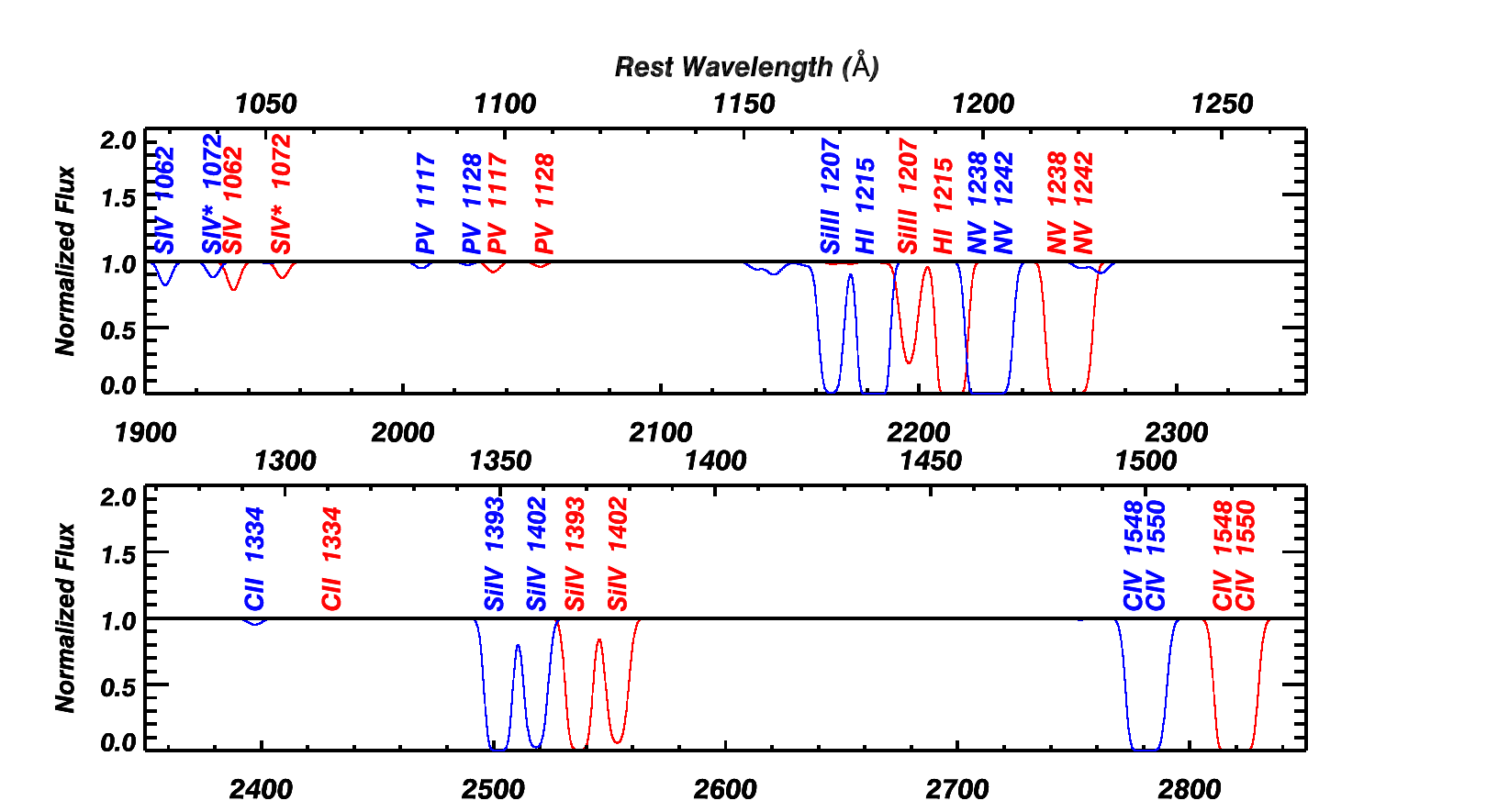}
\caption{The predictions of the strong absorption troughs for outflows S1 and S2 in the \ly\ $>$ 1050\angstrom\ rest frame from the best fitting photoionization models (see section \ref{sec:spec}). The models for outflow S1 and S2 are shown in blue and red, respectively.}

\label{fig:pred}
\end{figure*}

\subsection {Partial Covering and Ionization State Relationship}
\label{sec:PCIon}
Outflows are found to only partially cover the emission source \citep[e.g.,][]{Korista92, Arav99b, Arav01, Arav12, Hamann01}, and evidence exists to support the idea that the covering factor (f$_{\text{cov}}$) becomes larger when the level of ionization within the outflow increases. For example, \cite{Korista92} reported that the quasar 0226--1024  has an outflow with troughs from multiple doublet transitions arising from ions with different ionization potentials (IP). From the atomic data in \cite{Allen77}, IP(\ovi) = 77.41 eV $>$ IP(\civ) = 64.49 eV $>$ IP(\siiv) = 45.14 eV, and \cite{Korista92} found that the depths of the saturated absorption troughs from these doublets follow the same trend. Possible explanations include that the outflows have small dense cores covered by loose envelopes. Therefore, the low-density envelopes with larger f$_{\text{cov}}$ would tend to have higher-ionization levels than the high-density cores. However, elemental abundances are also found to affect the f$_{\text{cov}}$ \citep{Telfer98,Arav99a}. Studying different ionization states from the same element eliminates the abundance effects and provide us with a direct test of the relationship between f$_{\text{cov}}$ and ionization states. 

Outflow S2 shows absorption troughs from two different ions of oxygen: the \oiii\ (IP = 55.9 eV) multiplet around 834\angstrom\ and the \ovi\ (IP = 138.1 eV) doublet at 1031.93\angstrom\ and 1037.62\angstrom\ (see figure 1). From our photoionization solutions, the \HP\ produces a negligible amount of the \Nion\ for \oiii\ ($<$ 0.1\%) and almost ten times more \Nion\ for \ovi\ than the \LP. Therefore, \oiii\ and \ovi\ are good candidates for testing the difference in f$_{cov}$ between the phases. Our best fitting photoionization models predict that the absorption troughs from both of them are saturated, with N$_{model}$/N$_{mea}$ $\sim$ 5 and 30 for \oiii\ and \ovi, respectively (see table \ref{tb:IonSystems2}). However, the \oiii\ doublet shows non-black saturation with residual flux $\sim$ 30 -- 50\%, while the \ovi\ doublet has near zero residual flux (see figure \ref{fig:spec1}). This directly supports the idea that for the same element, higher-ionization ions indeed cover a larger area of the emission source. 

The \oiii\ and \ovi\ troughs in S1 show a similar behavior and support the same idea, as the saturated \oiii\ multiplet in S1 shows residual flux $\sim$ 10\% and the saturated \ovi\ doublet has nearly zero residual flux.




\subsection {The \ly\ $>$ 1050\angstrom\ Portion of the Outflow Spectra}
\label{sec:FUVpredict}

Ground-based BAL quasar outflow (BALQSO) studies mainly cover the rest frame wavelength range of \ly\ $>$ 1050\angstrom, which usually show absorption troughs from only \hi, \nv, \siiv, and \civ. 
The widest trough with a measurable width for S1 is the \oiii\ multiplet near 820\angstrom\ (rest frame) with $\Delta v$ = 2500 km s$^{-1}$ (measured for continuous absorption below the normalized flux I = 0.9). For S2,  we measure a width of 3100 km s$^{-1}$ from the \svi\ 933.38\angstrom\ trough. Therefore, both of these outflows are identified as broad absorption line outflows (BALs, see section 6.3 of Paper II for elaboration). 

From the best fitting photoionization solution derived in section \ref{sec:model}, we can predict the absorption features for the \ly\ $>$ 1050\angstrom\ rest frame region for each outflow by assuming this region has the same absorption trough shape as in the EUV500 region. In figure \ref{fig:pred}, we show the predicted troughs. For outflows S1 and S2, the predicted \civ\ \ly\ly 1548.19, 1550.77 absorption troughs are saturated, blended, and have widths of 2400 km s$^{-1}$ and 2600 km s$^{-1}$, respectively. Therefore, they are predicted to be BALs following the criteria of \cite{Weymann91}.


The models also predict weak absorption troughs for both \siv\ \ly
1062.66 and \siv*\ \ly 1072.97 with oscillator strengths ($f$) for both about 0.05,
which are the main density sensitive transitions for the \ly\ $>$
1050\angstrom\ rest frame region \citep{Arav18}. However, the
predicted troughs have maximum optical depths around 0.05, which make
their detection unlikely with ground-based telescopes for a couple of reasons. First,
$\tau=0.05$ troughs are difficult to detect in principle due to their
shallowness and systematic issues regarding the unabsorbed emission
model. Second and more important, from the ground we can detect the 
1062\angstrom\ rest-frame wavelength region only for quasars with redshifts
$z \gtrsim 2.5$. At these redshifts, the  \Lya\ forest severely contaminates the \siv\ troughs, 
which makes the task of identifying such shallow trough
hopeless in SDSS data and very difficult in  Very Large Telescope (VLT)/X-shooter
observations \citep[the latter observations have both higher S/N and
spectral resolution than the SDSS data, see][]{Arav18,Xu18a,Xu19}.

 In contrast, for the same outflow, we have four detected pairs of
\siv\ and \siv*\ troughs in the EUV500 with associated $f$ values up to
20 times larger (resulting in deeper troughs for the same amount of
\siv\ $N_{ion}$). Also the availability of two uncontaminated pairs of
\siv\ and \siv*\ troughs in S1, with large $f$ value differences, makes
the \ne\ determination more robust and less affected by possible
systematic issues. Based on these \siv\ and \siv*\ EUV500 troughs, we
were able to determine the \ne, $R$, and energetics for outflow
S1 (see section \ref{sec:distance}).

We also note that the predicted \pv\ troughs are even shallower than
the \siv\ \ly 1062.66 and \siv*\ \ly 1072.97 ones, which explains the
low detection rate of \pv\ troughs among BAL quasars \citep[3.0 -- 6.2\%, see][]{Capellupo17}.





\section{Summary}
\label{sec:summary}
In this paper, we analyzed outflows seen in the recent HST/COS spectra of quasar SDSS J0755+2306. The main results are summarized as follows:

1. Two outflow systems are identified. They present clear absorption troughs from both high-ionization species, e.g., \niii, \oiii, \oiv\ and \siv, and very high-ionization species, e.g., \arviii, \neviii, and \naix\ (see section \ref{sec:data}). Both outflows are classified as BALs from their widest EUV500 absorption trough widths.

2. Similar to the outflow analysis in Papers II, III, and V, each outflow system requires a two ionization-phase solution (see section \ref{sec:model}). 

3. For outflow system 2, we derive log(\ne) = 3.1 based on the density sensitive transitions of \oiii\ and \oiii* in the EUV500 band. The determined distance of this outflow is 1600 pc and the kinetic luminosity is $>$12\% of L$_{edd}$ (see section \ref{sec:distance} and \ref{sec:energy}). Therefore, this outflow is a good candidate for producing strong AGN feedback.

4. The absorption troughs from \oiii\ and \ovi\ support the idea that high-ionization ions have a larger covering fraction compared to lower-ionization ions (see section \ref{sec:PCIon}).

\acknowledgments
X.X., N.A., and T.M acknowledge support from NSF grant AST 1413319, as well
as NASA STScI grants GO 11686, 12022, 14242, 14054, 14176, and 14777, and NASA ADAP 48020.

Based on observations made with the NASA/ESA Hubble Space Telescope, obtained from the data archive at the Space Telescope Science Institute. STScI is operated by the Association of Universities for Research in Astronomy, Inc. under NASA contract NAS 5-26555.

CHIANTI is a collaborative project involving George Mason University, the University of Michigan (USA) and the University of Cambridge (UK).

\bibliography{apj-jour,dsr-refs}

\begin{thebibliography}{155}
\expandafter\ifx\csname natexlab\endcsname\relax\def\natexlab#1{#1}\fi

\bibitem[Allen(1977)]{Allen77} Allen, K.~W.\ 1977, Astrophysical quantities., by Allen, K.~W..~Translated from the 3.~revised and suppl.~English edition.~Moskva: Mir, 448 p., 


\bibitem[Arav et al.(1999a)]{Arav99a} Arav, N., Korista, K.~T., de Kool, M., Junkkarinen, V.~T., \& Begelman, M.~C.\ 1999, \apj, 516, 27 
\bibitem[Arav et al.(1999b)]{Arav99b} Arav N., Becker R.~H., Laurent-Muehleisen S.~A., Gregg M.~D., White R.~L., Brotherton M.~S., de Kool M., 1999, ApJ, 524, 566

\bibitem[Arav et al.(2001)]{Arav01} Arav, N., de Kool, M., Korista, K.~T., et al.\ 2001, \apj, 561, 118  
\bibitem[Arav et al.(2007)]{Arav07} Arav, N., Gabel, J.~R., Korista, K.~T., et al.\ 2007, \apj, 658, 829 
\bibitem[Arav et al.(2012)]{Arav12} Arav, N., Edmonds, D., Borguet, B., et al.\ 2012, \aap, 544, AA33
\bibitem[Arav et al.(2013)]{Arav13} Arav, N., Borguet, B., Chamberlain, C., Edmonds, D., \& Danforth, C.\ 2013, \mnras, 436, 3286
\bibitem[Arav et al.(2018)]{Arav18} Arav, N., Liu, G., Xu, X., et al.\ 2018, \apj, 857, 60 
\bibitem[Arav et al.(2019)]{Arav19} Arav, N., Xu, X., Kriss, G. A., et al.\ 2019, submitted to ApJ
\bibitem[Arav et al.(2020)]{ara20a} Arav, N., Xu, X., Miller, T.~R., et al.\ 2020, ApJS, in press



\bibitem[Bahk et al.(2019)]{Bahk19} Bahk, H., Woo, J.-H., \& Park, D.\ 2019, \apj, 875, 50 



\bibitem[Borguet et al.(2012a)]{Borguet12a} Borguet, B.~C.~J., Edmonds, D., Arav, N., Dunn, J., \& Kriss, G.~A.\ 2012a, \apj, 751, 107
\bibitem[Borguet et al.(2012b)]{Borguet12b} Borguet, B.~C.~J., Edmonds, D., Arav, N., Benn, C., \& Chamberlain, C.\ 2012, \apj, 758, 69 
\bibitem[Chamberlain et al.(2015)]{Chamberlain15b} Chamberlain, C., Arav, N., \& Benn, C.\ 2015, \mnras, 450, 1085 
\bibitem[Choi et al.(2014)]{Choi14} Choi, E., Naab, T., Ostriker, J.~P., Johansson, P.~H., \& Moster, B.~P.\ 2014, \mnras, 442, 440
\bibitem[Ciotti, Ostriker \& Proga(2009)]{Ciotti09} Ciotti, L., Ostriker, J.~P., \& Proga, D.\ 2009, \apj, 699, 89 
\bibitem[Capellupo et al.(2017)]{Capellupo17} Capellupo, D.~M., Hamann, F., Herbst, H., et al.\ 2017, \mnras, 469, 323 







\bibitem[Ferland et al.(2017)]{Ferland17} Ferland, G.~J., Chatzikos, M., Guzmn, F., et al.\ 2017, RMxAA, 53, 385





\bibitem[Gabel et al.(2006)]{Gabel06} Gabel, J.~R., Arav, N., \& Kim, T.-S.\ 2006, \apj, 646, 742 
\bibitem[Ganguly \& Brotherton(2008)]{Ganguly08} Ganguly, R., \& Brotherton, M.~S.\ 2008, \apj, 672, 102-107 

\bibitem[Gibson et al.(2009)]{Gibson09} Gibson, R.~R., Jiang, L., Brandt, W.~N., et al.\ 2009, \apj, 692, 758 
\bibitem[\protect\citeauthoryear{Green et al.}{2012}]{Green12} Green J.~C., et al., 2012, ApJ, 744, 60

\bibitem[Hamann et al.(2001)]{Hamann01} Hamann, F.~W., Barlow, T.~A., Chaffee, F.~C., Foltz, C.~B., \& Weymann, R.~J.\ 2001, \apj, 550, 142 

\bibitem[\protect\citeauthoryear{Hewett \& Foltz}{2003}]{Hewett03} Hewett P.~C., Foltz C.~B., 2003, AJ, 125, 1784

\bibitem[Hopkins \& Elvis(2010)]{Hopkins10} Hopkins, P.~F., \& Elvis, M.\ 2010, \mnras, 401, 7
\bibitem[Hopkins et al.(2016)]{Hopkins16} Hopkins, P.~F., Torrey, P., Faucher-Gigu{\`e}re, C.-A., Quataert, E., \& Murray, N.\ 2016, \mnras, 458, 816 




\bibitem[Korista et al.(1992)]{Korista92} Korista, K.~T., Weymann, R.~J., Morris, S.~L., et al.\ 1992, \apj, 401, 529 




\bibitem[Landi et al.(2013)]{Landi13} Landi, E., Young, P.~R., Dere, K.~P., Del Zanna, G., \& Mason, H.~E.\ 2013, \apj, 763, 86 



\bibitem[Miller et al.(2018)]{Miller18} Miller, T.~R., Arav, N., Xu, X., et al.\ 2018, \apj, 865, 90 
\bibitem[Miller et al.(2020a)]{mil20a} Miller, T.~R., Arav, N., Xu, X., et al.\ 2020, ApJS, in press
\bibitem[Miller et al.(2020b)]{mil20b} Miller, T.~R., Arav, N., Xu, X., et al.\ 2020, ApJS, in press
\bibitem[Miller et al.(2020c)]{mil20c} Miller, T.~R., Arav, N., Xu, X., et al.\ 2020, in preparation



\bibitem[Kramida et al.(2018)]{NIST18} Kramida, A., Ralchenko, Yu., Reader, J., and NIST ASD Team (2018). NIST Atomic Spectra Database (ver. 5.6.1), [Online]. Available: https://physics.nist.gov/asd [2019, April 26]. National Institute of Standards and Technology, Gaithersburg, MD. DOI: https://doi.org/10.18434/T4W30F



\bibitem[Ostriker et al.(2010)]{Ostriker10} Ostriker, J.~P., Choi, E., Ciotti, L., Novak, G.~S., \& Proga, D.\ 2010, \apj, 722, 642 


\bibitem[Prochaska et al.(2005)]{Prochaska05} Prochaska, J.~X., Herbert-Fort, S., \& Wolfe, A.~M.\ 2005, \apj, 635, 123 

\bibitem[Reichard et al.(2003)]{Reichard03} Reichard, T.~A., Richards, G.~T., Schneider, D.~P., et al.\ 2003, \aj, 125, 1711 


\bibitem[Scannapieco \& Oh(2004)]{Scannapieco04} Scannapieco, E., \& Oh, S.~P.\ 2004, \apj, 608, 62 
\bibitem[Schlafly \& Finkbeiner(2011)]{Schlafly12} Schlafly, E.~F., \& Finkbeiner, D.~P.\ 2011, \apj, 737, 103 


\bibitem[Trump et al.(2006)]{Trump06} Trump, J.~R., Hall, P.~B., Reichard, T.~A., et al.\ 2006, \apjs, 165, 1 
\bibitem[Telfer et al.(1998)]{Telfer98} Telfer, R.~C., Kriss, G.~A., Zheng, W., Davidsen, A.~F., \& Green, R.~F.\ 1998, \apj, 509, 132 
\bibitem[Tolea et al.(2002)]{Tolea02} Tolea, A., Krolik, J.~H., \& Tsvetanov, Z.\ 2002, \apjl, 578, L31 



\bibitem[Weymann et al.(1991)]{Weymann91} Weymann, R.~J., Morris, S.~L., Foltz, C.~B., \& Hewett, P.~C.\ 1991, \apj, 373, 23 
\bibitem[Wright(2006)]{Wright06} Wright, E.~L.\ 2006, \pasp, 118, 1711 




\bibitem[Xu et al.(2018)]{Xu18a} Xu, X., Arav, N., Miller, T., \& Benn, C.\ 2018, \apj, 858, 39 
\bibitem[Xu et al.(2019)]{Xu19} Xu X., Arav N., Miller T., Benn C., 2019, ApJ, 876, 105
\bibitem[Xu et al.(2020a)]{xu20a} Xu, X., Arav, N., \& Miller, T.\ 2020, ApJS, in press
\bibitem[Xu et al.(2020b)]{xu20b} Xu, X., Arav, N., \& Miller, T.\ 2020, ApJS, in press
\bibitem[Xu et al.(2020c)]{xu20c} Xu, X., Arav, N., \& Miller, T.\ 2020, ApJS, in press






 

 


%
%
 
%
%
%

\end{thebibliography}

\end{document}